\documentclass{aa}
\usepackage{txfonts}
\usepackage{natbib}
\usepackage{aalongtable}
\input epsf.sty
\def\sbz{S5~0716+714}
\begin{document}
\title{The Intra-Night Optical Variability of the bright BL Lac object 
S5~0716+714}
%
%
\author{F.~Montagni \inst{1}
        \and A.~Maselli \inst{2}
        \and E.~Massaro \inst{2}
        \and R.~Nesci \inst{2}
        \and S.~Sclavi \inst{2}
        \and M.~Maesano \inst{1}
        }

\offprints{enrico.massaro@uniroma1.it}

\institute{Stazione Astronomica di Vallinfreda, via del Tramonto, 
           I-00020 Vallinfreda, Italy
\and 
     Dipartimento di Fisica, Universit\`a La Sapienza, Piazzale A. Moro 2, 
     I-00185 Roma, Italy
           }

\date{Received ....; accepted ....}



\abstract
{}
{We address the topic of the Intra-Night Optical Variability of the BL Lac 
object \sbz.}
{To this purpose a long term observational campaign was performed, from 1996 
to 2003, which allowed the collection of a very large data set, containing 
10,675 photometric measurements obtained in 102 nights.}
{The source brightness varied in a range of about 2~mag, although the majority 
of  observations were performed when it was in the range $13.0 < R < 13.75$. 
Variability time scales were estimated from the rates of magnitude variation, 
which were found to have a distribution function well
fitted by an exponential law with a mean value of 0.027~mag/h,
corresponding to an e-folding time scale of the flux $\tau_F =$ 37.6~h.
The highest rates of magnitude variation were around 0.10--0.12 mag/h
and lasted less than 2~h. 
These rates were observed only when the source had an $R$ magnitude $<$ 13.4, 
but this finding cannot be considered significant because of the low
statistical occurrence.
The distribution of $\tau_F$ has a well defined modal value at 19~h.
Assuming the recent estimate of the beaming factor $\delta \sim$ 20,
we derived a typical size of the emitting region of about 
5$\times$10$^{16}/(1 + z)$ cm.
The possibility to search for a possible correlation between the mean
magnitude variation rate and the long term changes of the velocity of 
superluminal components in the jet is discussed.}
{}

\keywords{galaxies: active - galaxies: BL Lacertae objects -- 
          individual: S5~0716+714}

\authorrunning{F. Montagni et al.}
\titlerunning{The INOV of S5~0716+714}

\maketitle

\section{Introduction}  \label{sec:Intro}

\begin{figure*}
      \vspace{1.0cm}
      \hspace{2.0cm}
\epsfysize=9cm
\epsfbox{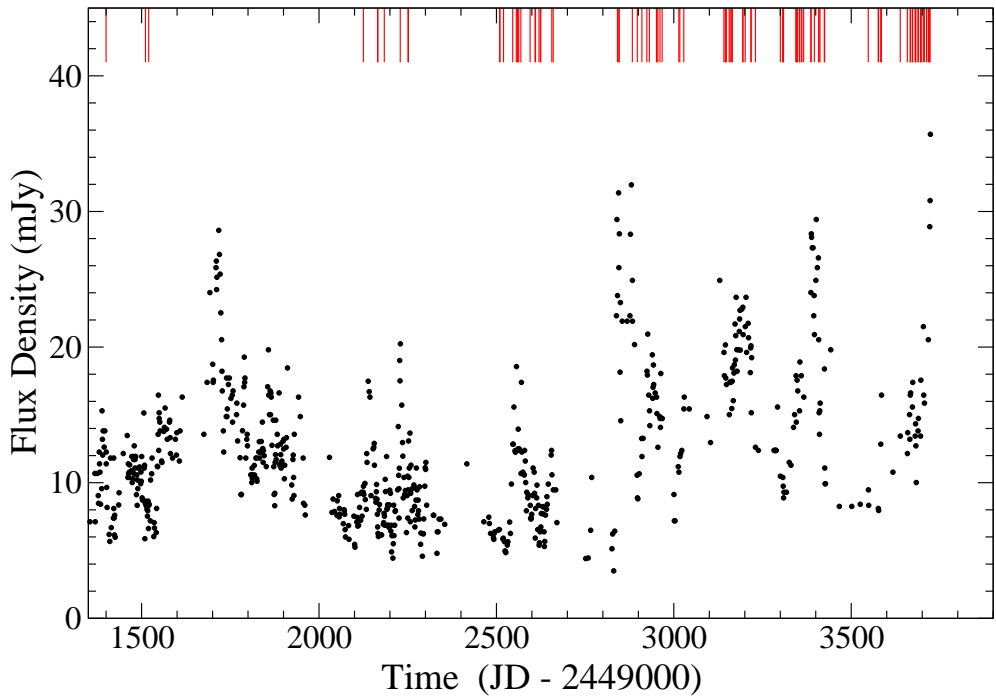}
\caption[]{
The light curve in the $R$(Cousins) band of \sbz~ from February 1997 to 
March 2003. Vertical bars on the top mark the epochs of our INOV
observations.  
}
\label{fig:lc0716}
\end{figure*}

The radio source \sbz~ was identified with a bright and highly 
variable BL Lac object, characterized by a strong  featureless
optical continuum (\citealt{Bie81}). 
The failure in detecting a host galaxy both in HST direct imaging 
(\citealt{Urr00}) and in high S/N spectra (\citealt{RecSto01}) 
suggests that its redshift $z$ should be greater than 0.52 
(\citealt{Sba05}; see also \citealt{Sch92}  and \citealt{Wag96}).
Variations on short time scales (a fraction of hour) have been
detected in several occasions at frequencies ranging from radio to X-rays 
(\citealt{Qui91}; \citealt{Wag96}; \citealt{Gab00}; \citealt{HeiWag96};
\citealt{Nes02}; \citealt{Gio99}; \citealt{Vil00}; \citealt{Wu05}).
The recent history of the flux of \sbz~ in the $R$ (Cousins) band is plotted 
in Fig.~\ref{fig:lc0716}, which spans the time interval from 1997 to 2003. 
The photometric points up to 2001 have been extracted from the data base by 
\citet{Rai03},
taking only one measurement per day, whereas points from 2002 to 2003 
are new data obtained by our group (\citealt{Nes05}).
The general structure of this curve shows that the mean flux of \sbz~
varies in the range 5-20~mJy with some flares in which it reaches 
and overwhelms 30~mJy.
These flares are separated by time intervals variable from 
$\sim$~1 to $\sim$~3 years.
In Fig.~\ref{fig:lc0716}  there are four large flares having a typical 
duration of the order of 1--2~months from which it is possible 
to estimate a flaring duty cycle of about 5--10\%.

In this paper we focus our attention on the so called Intra-Night
Optical Variability (INOV) or {\it microvariability}. 
Brightness changes of BL Lac objects, having amplitudes of about 10--20\% 
and occurring on time scales as short as a fraction of an hour, were studied 
since the eighties by Miller and coworkers 
(\citealt{Mil89}; \citealt{Car91}).
This phenomenon was after detected in many sources of this class and
it could be considered one of their characterizing properties.
INOV in BL Lac objects and other blazars has been investigated by 
several authors, using either single or multi band photometry: 
among BL Lac objects showing such activity we recall 
AO 0235+164 (\citealt{Rom00}),
S4 0954+658 (\citealt{Pap04}),
BL Lacertae (\citealt{Mas98}; \citealt{Nes98}; \citealt{Pap03})
whereas studies on samples of sources are those on LBL (Low energy peaked 
BL Lacs) objects (\citealt{HeiWag96}),
EGRET blazars (\citealt{Rom02}) 
and other BL Lac objects and radio-core dominated blazars 
(\citealt{Sag04}; \citealt{Sta05}).
These works are generally based on data sets, for single sources,
obtained in a not high number of nights and/or observations. 
Moreover, the analysis is mainly focused on the search for recurrence 
time scales detectable in the individual light curves.

A first detailed study of INOV in \sbz~ is that of 
\citet{Wag96},
who investigated rapid variations in multifrequency (from radio to 
X-rays) campaigns and observed a quasi-periodic behaviour with a typical 
recurrence time of about two days and a high correlation between the 
optical and radio flux changes.
The possibility of an harmonic component in the optical flux of \sbz~ was 
after confirmed by 
\citet{HeiWag96},
who did not detect a similar effect on the other BL Lacs of their sample.
\citet{Sag99}
reported INOV multiband observations covering 4 weeks in 1994 and found 
only three major events of rapid variability in which the highest 
magnitude variation rate was around 0.03~mag/h. 
\citet{Vil00}
reported the results of a WEBT (Whole Earth Blazar Telescope) campaign 
on \sbz~ from 16 to 22 February 1999 in which a relevant INOV was observed 
every night with magnitude variation rates up to $\sim$~0.1~mag/h.
 
A well established definition of INOV properties for a given source 
(or for a class of sources) can be obtained only from the analysis of 
a relatively large number of observations, possibly performed in different 
brightness states. 

The definition of time scale for non-periodic phenomena is not
univocal and can depend on the type of variation: in this work we consider 
 a time scale based on the magnitude variation rates. 

We worked extensively on \sbz, bright enough to obtain good photometric data 
with small aperture telescopes and short exposure times. 
Our INOV observational campaign of \sbz~ started in 1996 and since 
November 1998 we undertook a more intense data acquisition concluded 
in spring 2003. 
In this paper we report a large set of observational data, containing 
10,675 photometric points, obtained in 102 nights, which is up to now the 
largest database of INOV for any BL Lac object. 
Our statistical analysis will give new informations on the distribution 
of the variability time scales and other properties of this source.

\section{Observations and Data Reduction}  \label{sec:ObsDR}
Most of our photometric observations of \sbz~ were performed with a 
32~cm f/4.5 Newtonian reflector located near Greve in Chianti (Tuscany) 
with a CCD camera, manufactured by DTA, mounting a back-illuminated 
SITe SIA502A chip.  A few observations were made with the 50~cm telescope 
of the Astronomical Station of Vallinfreda (Rome) and the 70~cm telescope, 
formerly at Monte Porzio (Rome), both equipped with CCD cameras.
Standard  $B$, $V$ (Johnson), $R$ (Cousins) and $I$ (Cousins) filters 
were used. Exposure times depended upon the brightness of the source and 
varied between 3 and 5 minutes to have a typical noise level on the 
comparison stars of 0.01~mag or less.

Differential photometry with respect to three or four comparison stars
in the same field of view of \sbz~ was performed using the {\it apphot}
task in IRAF 
\footnote{Image Reduction and Analysis Facility, distributed 
by NOAO, operated by AURA, Inc. under agreement with the US NSF.}.
The same circular aperture, with a radius of 5~arcsec was used for the 
photometry of \sbz~ and the comparison stars.
These were the A, B, C, D stars of the reference sequence given by
\citet{Ghi97},
corresponding to stars 2, 3, 5, 6 of the sequence by 
\citet{Vil98}.
Our large number of images gave the possibility to calculate (for our 
bandpasses and detector response) their magnitude intercalibration with a high 
accuracy and consequently we modified the original $B$, $V$, $R$  values to 
minimize the measured uncertainties, while for the $I$ band those given by 
\citet{Ghi97}
were unchanged. The adopted values are listed in Table~\ref{tab:BVRImag}. 

\begin{table}
\caption{ $B$, $V$, $R$, $I$ magnitudes of reference stars used in our 
analysis.}
\label{tab:BVRImag}
\centering
\begin{tabular}{ccccc}
\hline\hline
Star & $B$ & $V$ & $R$ & $I$ \\
\hline
A, 2 & 12.03 & 11.51 & 11.20 & 10.92 \\
B, 3 & 13.06 & 12.48 & 12.10 & 11.79 \\
C, 5 & 14.17 & 13.58 & 13.20 & 12.85 \\
D, 6 & 14.25 & 13.66 & 13.28 & 12.97 \\
\hline
\end{tabular}
\end{table}

We took as best estimate of the errors of the \sbz~ magnitudes the rms value 
of the reference stars' values combined with the statistical error of the 
pixel counts. These resulted generally of the order of 0.01~mag.
We verified {\it a posteriori} that the typical scatter of the data of 
\sbz~ around a smoothed 4-6 point running average curve was fully 
compatible with this error estimate.
In a small number of cases we found some data showing large discrepancies
with respect to the nearest points or having uncertainties much larger 
than the others. We believed that such large differences are of instrumental 
origin, due to a possible occurrence of hot/cold pixels or a noise fluctuation.
We preferred to cancel these data from the light curves rather than to correct 
their values: their number was generally quite small and light curves were
unaffected by the removal.
Moreover, some light curves were taken in unstable and not photometric nights
and the data scatter with respect to the running averages was higher.
Noisy segments or entire curves were discarded from the data set
to avoid the inclusion of poor data and possible spurious effects. 

\begin{figure}
      \vspace{-1.0cm}
\epsfysize=9.0cm
\epsfbox{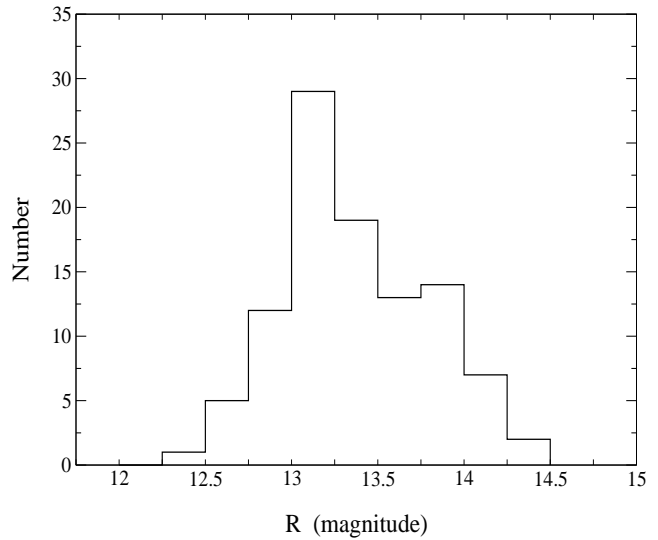}
\caption[]{
Histogram of the mean $R$ magnitude in the nights of INOV observations.
}
\label{fig:histRmean}
\end{figure}

Table~\ref{tab:logobs} reports the log of all good quality observations 
considered in the present paper: 
for each night we give the date of observation (column 1), the abridged JD 
(column 2), the UT start time (column 3), the observation duration (column 4),
the number of frames (column 5), the filter used (column 6), the mean 
magnitude of the source in the used filter 
(column 7), the rms deviation of the magnitude difference between two 
reference stars (column 8), the telescope (column 9: G=Greve, M=Monte Porzio, 
V=Vallinfreda) and the number of time intervals used in the evaluation of the 
magnitude variation rates (column 10).
The entire photometric data set including measured magnitudes of \sbz, 
not corrected for the interstellar reddening, and errors is given in
Table~A1, described in the Appendix and available in electronic form at
CDS.

An histogram of the mean magnitude of \sbz~ during our intranight
observations is given in Fig.~\ref{fig:histRmean}. 
The source brightness varied in a range of about 2~mag, but the majority 
of observations were performed when it was in the range $13.0 < R < 13.75$.

As the host galaxy is undetected (an upper limit of $R > 20$~mag 
is given by \citealt{Urr00})
no correction needs to be applied to our photometric values.

\begin{figure*}
      \vspace{1.0cm}
      \hspace{3.0cm}
\epsfysize=16cm
\epsfbox{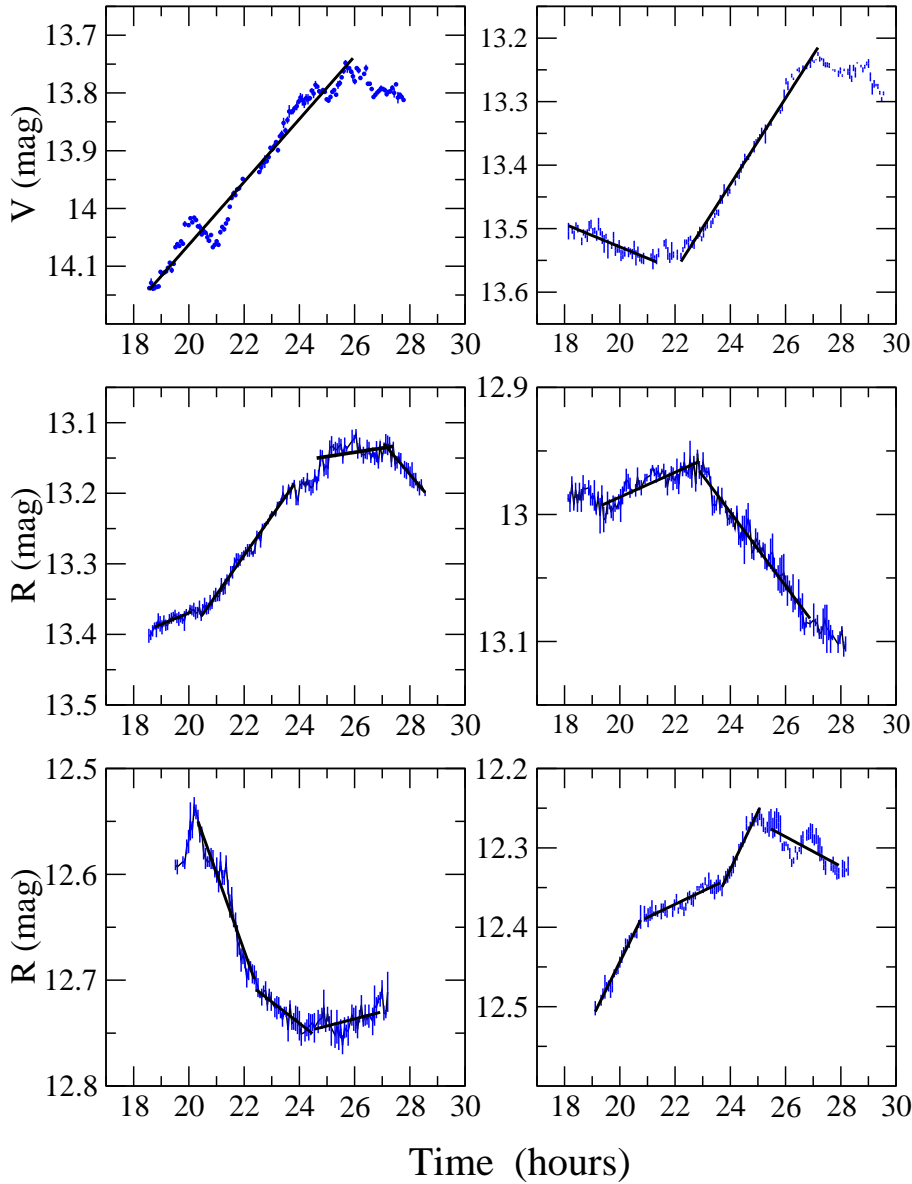}
\caption[]{
Some examples of INOV of \sbz~ observed at different epochs from 2000
to 2003. The solid lines through the data show the linear best fit  
used to evaluate $|\Delta m/\Delta t|$  in each selected interval. 
Date of observations are 02-01-2000 (top left), 12-01-2000 (top right), 
26-02-2001 (middle left), 03-11-2001 (middle right), 22-04-2002 (bottom left), 
23-03-2003 (bottom right). 
}
\label{fig:exINOV}
\end{figure*}

\section{Light curves and variability time scale} \label{sec:LC}
As mentioned in the Introduction, INOV is a frequent characteristic 
of the optical emission of \sbz.
We detected significant brightness changes in a large fraction of nights:  
only in 9 nights (marked with an asterisk in Table~\ref{tab:logobs}) 
the source remained stable during the observation time window.
Some light curves showing examples of relevant INOV of \sbz~ are plotted
in the panels of Fig.~\ref{fig:exINOV}: in these cases the variation 
amplitudes were larger than 0.1~mag and overwhelmed 0.3~mag in some of them. 
In some occasions light curves show oscillations of 
about 0.05--0.1~mag with a typical duration of a few hours. 
From Table~\ref{tab:logobs} we see that the rms value of the difference 
between two reference stars was generally within one or two hundredths of 
magnitude confirming the reality of these oscillations. Some examples will be
described in Sect.~\ref{sec:Microosc}.

The time scale of flux variations can be defined as
\begin{equation}    \label{eq:tauf}
\tau_F = \frac{1}{1 + z}\,\,\frac{F_{\nu}}{|dF_{\nu}/dt|}
\end{equation}
which, when it can be considered constant, corresponds to the e-folding 
time scale. The factor $1 + z$ takes into account cosmological effects.
This definition of variability time scale is similar, but not coincident,
with that used by  \citet{Rom02},
who adopted the flux variation $\Delta F$ measured in the night instead 
of~$F$. With their definition, time scales resulted comparable or shorter 
than the durations of the observations, while with our definition the result 
is less related to the length of time windows. 
The value of $\tau_F$ can be directly computed from the magnitude light curves 
$m(t)$, because its variation rate corresponds to the log-derivative of~$F$. 
In fact, from the classic Pogson's formula of astronomical photometry,  
we have: 
\begin{equation} \label{eq:dmdt}
\bigg|\frac{dm}{dt}\bigg| = 2.5~\bigg|\frac{d}{dt}Log~F_{\nu}\bigg| = 
\frac{1.086}{\tau_F}
\end{equation}

\begin{figure}
\vspace{-1.0cm}
\epsfysize=9.0cm
\epsfbox{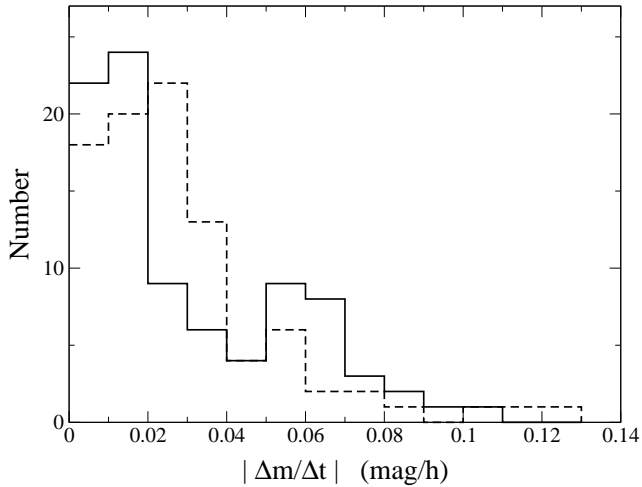}
\caption[]{
The solid histogram is the number distribution of the decreasing values 
of $\Delta m/\Delta t$ while the dashed one is that for the increasing values.
}
\label{fig:histKS}
\end{figure}

To derive the values of $\tau_F$ we divided each light curve into a few 
monotonic intervals. 
 For each interval we fitted the light curve with a straight line and
accepted the best fitting slope value as our estimate of $\Delta m/\Delta t$.
Obviously,
the selection of these segments is a delicate task that cannot be performed 
by means of a simple algorithm because of the large variety of light 
curve shapes. 
We were aware that subjective selection criteria, which can introduce some 
bias, are unavoidable and in order to reduce this possibility we adopted the 
following procedure:
\begin{itemize}
\item
the number of selected intervals per night must be as small as possible;
\item
the selected time intervals must be long enough to contain a sufficient number 
($> 10$, but typically $> 20$, i.e. two hours) of data points; 
\item
 time intervals cannot overlap;
\item
segments must be compatible with a constant magnitude variation rate, 
which was verified looking for trends in the residuals with respect to 
a linear best fit;
\item
magnitude oscillations of small amplitude and short duration superposed to 
a longer trend, like those described below in Sect.~\ref{sec:Microosc}, 
were not considered to evaluate $\Delta m/\Delta t$.
\end{itemize}

In all cases a linear fit was fully adequate to describe the selected 
segments of the light curves with no systematic structure left in the 
residuals.

The number of time intervals taken into account for each night is 
indicated in Table~\ref{tab:logobs}.
The total number of intervals is 204 with a mean number per night 
equal to 2. 
Only in 8 nights, over a total number of 102, light curves were found
particularly structured to take into account 4 or 5 intervals. 
Note, however, that 5 of these nights are in the last month of our campaign 
(from the end of February to that of March 2003), that could correspond to 
a period in which \sbz~ exhibited variations faster than usual. 
Finally we recall that only in 9 nights the source was stable (formal fit
slope less than 0.002~mag/h).

We checked if there were differences in the distributions of the rising 
and decreasing values of $\Delta m/\Delta t$. 
A Kolmogorov-Smirnov test gives no difference at the 98.3\% confidence level: 
the histogram is reported in Fig.~\ref{fig:histKS}.
Therefore the statistical analysis of the distribution of $\Delta m/\Delta t$
(and corresponding $\tau_F$) is performed on their absolute values.

\begin{figure}
\vspace{-1.0cm}
\epsfysize=9.0cm
\epsfbox{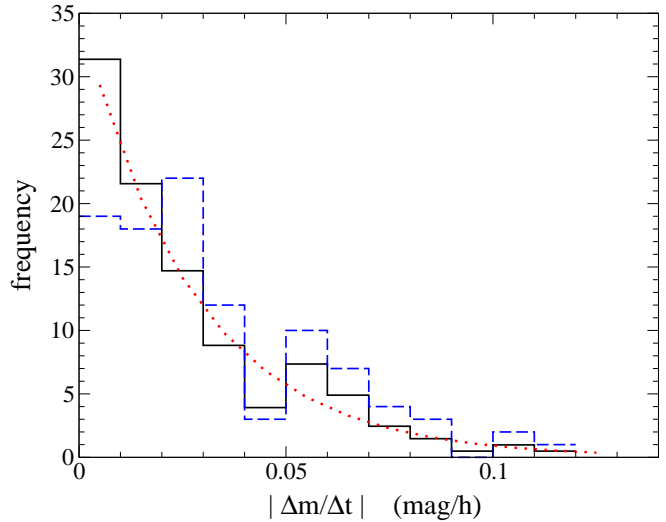}
\caption[]{
The solid histogram is the frequency distribution of all the variation rates 
$|\Delta m/\Delta t|$ (in mag/h) in all the used filters. 
Dotted line is the exponential law best fit.
The dashed histogram is that  obtained using just the highest rate 
($|\Delta m/\Delta t|_{max}$) measured in each night.
}
\label{fig:histdist}
\end{figure}

Fig.~\ref{fig:histdist} shows the frequency histogram 
(i.e. number of data in each bin divided by the bin width)
of all measured $|\Delta m/\Delta t|$ that can be very well represented
by the exponential distribution  (dashed line):
\begin{equation}  \label{eq:fabsmdot}
 f(|\dot{m}|) = K~exp(-|\dot{m}|/\langle|\dot{m}|\rangle)
 \end{equation}
 with $\langle|\dot{m}|\rangle$ = 0.027~mag/h and the best fit value of the
 normalisation constant $K =$ 36.5, very close to the theorical value
 $1/\langle|\dot{m}|\rangle$.

\begin{figure}
      \vspace{-1.0cm}
\epsfysize=9.0cm
\epsfbox{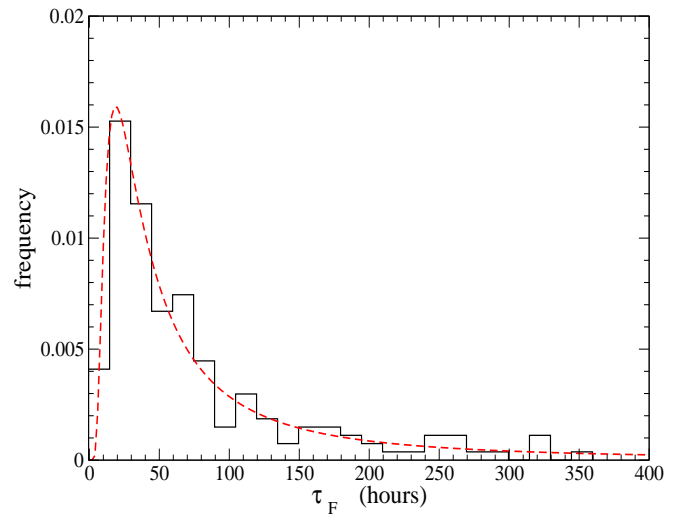}
\caption[]{ Frequency histogram of the variability time scales $\tau_F$. 
Dashed line is the best fit of the theorical distribution given by 
Eq.~(\ref{eq:ftauf}).
}
\label{fig:histvartau}
\end{figure}

The statistical distribution of $\dot{m}$ (the time derivative of $m$, 
in our case $|\Delta m/\Delta t|$)
and of $\tau_F$ satisfy the relation:
\begin{equation}  \label{eq:fmdot}
 f(\dot{m}) = \frac{1}{\tau_F^2} f(\tau_F) 
\end{equation}

 Appling this relation, the distribution of $\tau_F$ is
\begin{equation}  \label{eq:ftauf}
 f(\tau_F) = \frac{K'}{\tau_F^2}~exp(-\tau_F^*/\tau_F) 
\end{equation} 
with $\tau_F^* = 1.086/\langle |\dot{m}| \rangle $.
Fig.~\ref{fig:histvartau} shows the resulting histogram for $\tau_F$ and 
the best fit of the distribution given in Eq.~(\ref{eq:ftauf});
in particular, $\tau_F^*$ is
equal to 37.6~h, in a very good agreement with the expected value.

In computing the parameters of this distribution we did not include very small 
$|\dot{m}|$ values which would give $\tau_F$ longer than 350~h.
We point out that the mean value of variable $\tau_F$ cannot be computed 
from the first moment of Eq.~(\ref{eq:ftauf}) because, once multiplied by 
$\tau_F$, the integral does not converge. 
So a direct calculation of $\langle \tau_F \rangle$ from the values 
is not statistically correct: indeed a very small value of $|\dot{m}|$ 
(a flat light curve segment) corresponds to a very high $\tau_F$ 
that pushes the mean towards an infinit value. A much better statistically 
defined parameter useful to describe this distribution is the mode 
$\tau_{Fm}$. It can be calculated from the maximum of Eq.~(\ref{eq:ftauf}), 
which occurs at $\tau_{Fm} = \tau_F^*/2$, corresponding to 18.8~h.
 
\begin{figure}
      \vspace{-1.0cm}
\epsfysize=9.0cm
\epsfbox{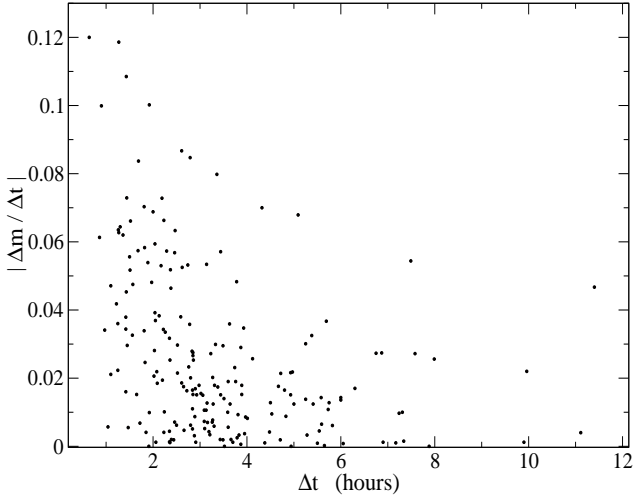}
\caption[]{
 Plot of $|\Delta m/\Delta t|$ {\it vs} $\Delta t$, the magnitude 
variation rates against the duration of the corresponding time intervals. 
}
\label{fig:bivplot}
\end{figure}

In Fig.~\ref{fig:histdist} we plotted also the histogram (dashed line)
of $|\Delta m/\Delta t|_{max}$ obtained considering the highest variation 
rates measured in each night: the content of the bins above 0.03~mag/h is 
slightly changed, whereas the frequencies in two lower bins are much smaller.
This histogram is useful to evaluate the probability of 
how large a magnitude variation rate can be
in the typical observation window of a night (say eight-nine hours), even when 
the time interval in which the variation occurs is shorter.
We see that rates $\sim$ 0.03~mag/h are the most frequently measured.

We also studied whether there is a relation between $|\Delta m/\Delta t|$ and 
the duration of the time intervals $\Delta t$. 
Fig.~\ref{fig:bivplot} shows the  plot of these two variables: we can see 
that most intervals have durations in the range 2--4 h and that in only few 
cases durations longer than 9~h were detected. 
Note that in three of the latter occasions we measured low rates 
and only in one case it was close to 0.05~mag/h. 
Variation rates greater than 0.1~mag/h were measured only over time intervals
shorter than 2~h, indicating that such high rates do not remain stable for  
long time intervals. 
On the other side, note that rates smaller than 0.02~mag/h were found for
intervals of any duration confirming that the probability to observe the 
source brightness steady is much higher than that to find a large INOV, 
as expected from the exponential distribution in Fig.~\ref{fig:histdist}.
We can also conclude that magnitude variation rates higher than 0.1~mag/h 
are a rare phenomenon having a probability lower than a few percent. 

\begin{figure}
      \vspace{-1.0cm}
\epsfysize=9.0cm
\epsfbox{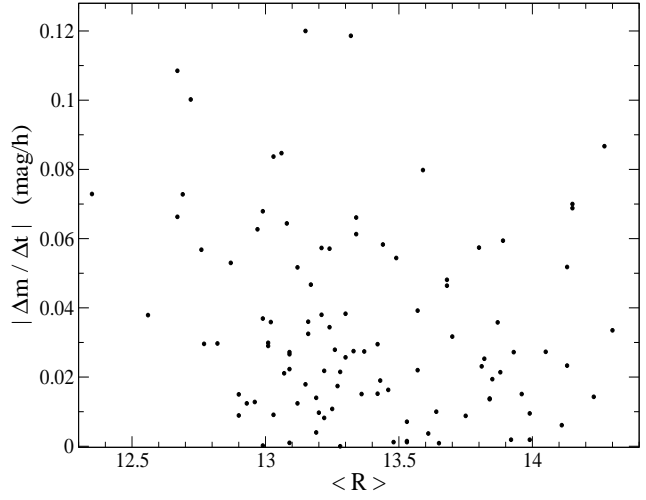}
\caption[]{
The variation rates (in mag/h) plotted against the mean $R$ magnitude 
of the source. Note that the rates above 0.085~mag/h were found only when 
\sbz~ was brighter than $ R = 13.4$, while lower rates were found in any 
brightness states.
}
\label{fig:varratmag}
\end{figure}

We also searched for a possible relation between INOV and the mean brightness
state of \sbz. 
The plot in Fig.~\ref{fig:varratmag} shows the distribution of largest values 
of $|\Delta m / \Delta t|$ measured in each light curve against the mean 
value of $R$ in that night.
When the used filter was different we derived approximate $R$ value applying
the mean colour indices given by \citet{Rai03}, 
which were found generally stable.
The points corresponding to rates with $|\Delta m / \Delta t| <$ 
0.08~mag/h appears to be rather uniformly distributed, whereas a possible 
deviation from uniformity can be seen in the upper region, corresponding 
to the highest rates. 
These rates, in fact, were measured only when \sbz~ was brighter than 
$R \simeq$~13.4, an indication that a faster INOV is more likely 
when the source is brighter: the statistical significance of this finding, 
however, is limited by the small number of measures.
We remark, anyway, that these high rates were not found within a short 
observation window, but are distributed from 2000 to 2003.

Finally we briefly comment on the possibility that the magnitude 
variation rates can be different for different photometric bands. 
Actually we have 3 nights in B band, 6 in I, 15 in V and 78 in R 
(see Table \ref{tab:logobs}:
the number of nights with filters different from R is very limited, and the 
variation of the color index of the source in the historical data base
(\citealt{Rai03})
is rather small, and without a clear correlation with the source luminosity, 
so we do not expect a detectable difference in the distribution of the 
$\tau_F$ values. 
A formal Kolmogorov-Smirnov test between the $\tau_F$ distribution derived 
from the R-band data and that derived from the BVI ones gives no indication 
of differences.

\section{Micro-oscillations} \label{sec:Microosc}
In the previous section we mentioned that in several occasions the 
INOV of \sbz~ was characterized by oscillations of very small amplitude, 
typically $<$~0.10 mag. 
Two examples of such a behaviour can be seen in Fig.~\ref{fig:exINOV} and 
precisely in the light curves of 2000 January 2 (top left panel) and 
2003 March 23 (bottom right panel).
Recently
\citet{Wu05}
reported the observation of a similar oscillation with an amplitude
of 0.05~mag, a duration of about 5 hours and without a detectable 
colour change. 
They discuss also about the possible origin of this type of INOV 
and conclude that it could be due to a geometric modulation as 
expected in the helical jet model by 
\citet{CamKro92}.

\begin{figure}
      \vspace{0.0cm}
\epsfysize=9.0cm
\epsfbox{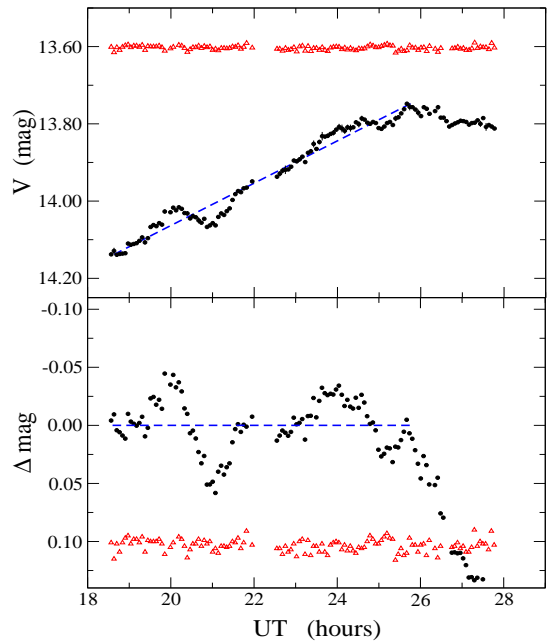}
\caption[]{
The light curve in the $V$ band observed on 2000 January 2.
Upper panel shows photometric data of \sbz~ (filled circles) and the magnitude
difference of two reference stars (open triangles), shifted by an arbitrary
offset; the dashed line is the linear best fit of a brightening segment.
Lower panel shows the difference of \sbz~ magnitude with respect
to the best fit line to show the micro-oscillations: in this case the
amplitude is $\sim$~0.05~mag and the duration of a cycle is $\sim$~3~h.
}
\label{fig:lc02Jan00}
\end{figure}

Small amplitude oscillations are occasionally superposed onto variations
characterised by longer trends. 
Fig.~\ref{fig:lc02Jan00} shows the light curve observed on 2000 January 2 
(see also Fig.~\ref{fig:exINOV}, top left panel) when a couple of consecutive 
micro-oscillations were detected in the interval from 18.60 to 25.85~UT. 
To make these oscillations more evident we fitted the interval where they are 
present with a straight line and subtracted the interpolated magnitudes 
from the original data set.
The curve of the magnitude differences with respect to the linear best 
fit is plotted in the lower panel: the resulting oscillation amplitude 
is $\sim$~0.05 mag and the duration is about 3~h.
For comparison we plotted in both panels the differences of the magnitudes
of two reference stars: they remain much more stable than the source, not 
only in the magnitude variation shown in the upper panel, but also with 
respect to the differences in the lower panel. 
Note also the regularity of these changes that cannot be confused
with a spurious effect due to noisy observations. 
We conclude that this particular INOV can be originated neither by
variations of the atmospheric extinction nor by other instrumental effects
and that it must be considered genuine.

\begin{figure}
\vspace{-1.0cm}
\epsfysize=8.0cm
\epsfbox{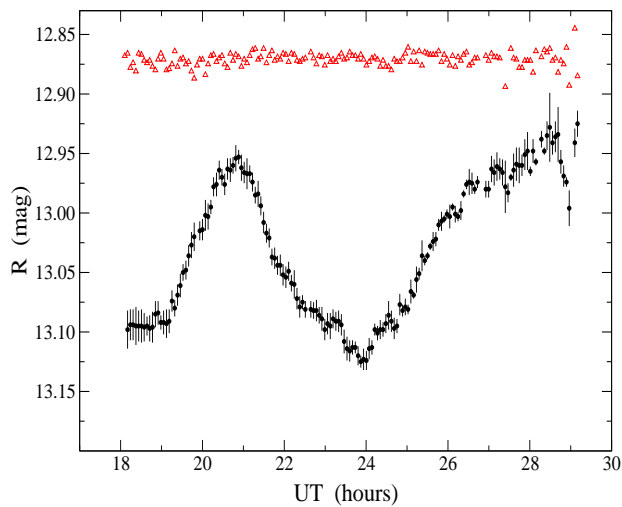}
\caption[]{
Micro-oscillation of the brightness of \sbz~ observed on 2003 February 25.
Triangles are the difference between the magnitude of two reference stars
with an arbitrary offset. Note the noise increase in the last two hours
in both light curves.
}
\label{fig:lc25Feb03}
\end{figure}

Fig.~\ref{fig:lc25Feb03} shows another micro-oscillation detected on
2003 February 25.
In this case the amplitude was of $\sim$~0.1~mag with a duration of
$\sim$~6.5~h.
Because of its relatively long duration it was segmented into three
intervals to evaluate the magnitude variation rate.
Note in the last portion of the curves of both \sbz~ and reference stars
the effect of a noise increase likely due to the incoming twilight.

Note also that the light curve of 2000 January 12 (see the upper right panel 
in Fig.~\ref{fig:exINOV}) could be considered as a micro oscillation with 
respect to the mean brightening trend covering all the night. 
In this case it would have an amplitude of $\sim$~0.1~mag and a
duration of $\sim$~11~h.

The oscillating behaviour was already noticed in other occasions 
(\citealt{Wag96}; \citealt{HeiWag96}; \citealt{Vil00}).
All these examples indicate that the micro-oscillating behaviour is 
not particularly rare but, at the same time, it 
 does not have a stable pattern. 
A quantitative estimate of their frequency, however, is rather difficult, 
because their detection is limited by the duration of the observing window.
The examples presented above suggest the possibility that longer  
variations can have  typically larger amplitudes than the short ones. 
The confirmation of this hypothesis on a statistical ground 
requires a collection of INOV light curves even greater than ours.

\section{Discussion} \label{sec:Disc}
The observational results described in this paper are useful to
extend the present knowledge about INOV of BL Lac objects.
In an about six year long campaign we obtained a large collection of data
on \sbz, not available before for a single source, useful to develop
a statistical study of the main INOV properties.
We give in Table~A1 (see Appendix) the whole data set, including all 
10,675 photometric measurements. They can be used for further investigations 
and for a comparison with other data on the same and other sources.

Our statistical analysis was essentially based on the evaluation of 
magnitude variation rates $|\Delta m/\Delta t|$ over several time intervals, 
selected using rather uniform criteria to minimize possible biases.
We found that the resulting distribution is fully compatible with 
an exponential one having a mean 
$\langle |\Delta m/\Delta t| \rangle =$ 0.027~mag/h, 
corresponding to a flux variation time scale of~37.6~h. 
This finding implies that the probability to observe a magnitude variation 
rate higher than 0.2~mag/h is smaller than 10$^{-3}$, and therefore one 
would require more than 500 nights of observations like ours to detect 
an episode having such a high rate.

The interpretation of the variability of blazars is not a simple problem 
because it involves the description of rapidly changing processes 
characterized by several physical quantities, whose mean values and
statistical distributions are poorly known.
The fact that we found an exponential distribution for 
$|\Delta m/\Delta t|$ without any evidence of a typical time scale suggests 
that the INOV is essentially a stochastic process.
A possibility already considered in some papers is that of a turbulent jet.
A model in which relativistic electrons emit synchrotron radiation in a
turbulent magnetic field was described by \citet{Mar92}:
the resulting light curves show trends and oscillations like those described
in the previous sections. This agreement is, however, only qualitative
and a larger observational effort should be performed to achieve a more
detailed description of the turbulence parameters. 
For instance, the discovery of a relation between the amplitude and the 
duration of small oscillations on a robust statistical ground can help to 
 model the turbulence. 

Another investigation can concern the possible long term variations of INOV 
parameters.
\citet{Nes05} 
recently presented the historic light curve of \sbz~ that shows a well 
apparent brightening trend since about 25 years. 
A study of the apparent ejection velocities of superluminal blobs in 
the jet $\beta_{ej}$ (\citealt{Bac05})
showed that it decreased from $\sim$~15 to $\sim$~5 in the period from 
1986 to 1997. 
Both effects are consistent with a scenario of a precessing jet having a 
stable Lorentz factor $\Gamma \simeq 12$  but approaching the line of 
sight from an angle of about 5$^{\circ}$ to 0$^{\circ}$.5. 
The corresponding Doppler factor $\delta = 1/\Gamma (1-\beta\,cos\theta)$ 
($\Gamma$ is the bulk Lorentz factor, and $\theta$ the angle between the
jet direction and the line of sight) increased from $\sim$~13 to $\sim$~25.

We can use this independent estimate of $\delta$ to constrain the size of the
emitting region responsible for INOV.
Considering the mode $\tau_{Fm}$ of the distribution in 
Fig.~\ref{fig:histvartau} as a typical (short) timescale and assuming 
$\delta \simeq 20$ 
(\citealt{Nes05}),
we can derive an upper limit to the characteristic size of the emitting 
region in the comoving frame
$r' \simeq c \,\delta\, \tau_{Fm}/(1 + z) \simeq 
5\times10^{16}/(1 + z)$ cm, a value which agrees well with those usually 
adopted in modelling BL Lac jets.

The distribution of the magnitude variation rates is useful to detect a 
change in the jet orientation. Indeed, if synchrotron emission is originated 
in a relativistic jet, the observed flux is related to the one emitted in the 
comoving frame~$F'$ (assumed practically steady) by the relativistic boosting:
\begin{equation}   \label{eq:ft}
  F(t) = (\delta(t))^k F' 
\end{equation}
where $k = 3 + \alpha$ ($\alpha$ is the spectral index). 
After converting the flux in magnitude and deriving with respect to the time:
\begin{equation}  \label{eq:absmdotdelta}
|\dot{m}| = 1.086~k\,|\dot{\delta}/\delta| 
\end{equation}
If the variation of $\delta$ is due only to the change of $\theta$, we obtain
\begin{equation}  \label{eq:absdeltadotdelta}
 |\dot{\delta}/\delta| = \beta_{app} |\dot{\theta}| 
\end{equation}
and 
\begin{equation} \label{eq:absmdotbeta}
 |\dot{m}| = 1.086\,k\,\beta_{app} |\dot{\theta}| 
\end{equation}
where $\beta_{app} =(\beta~sin \theta)/(1 - \beta~cos\theta) = - d 
(1 - \beta~cos\theta)^{-1}/d\theta$ is the apparent velocity of superluminal
components along the jet.
This relation suggests that in the case of a regular variation 
of $\theta$  (precessing jet) it is possible to expect
a positive correlation between $\langle|\dot{m}|\rangle$ and $\beta_{app}$. 
Under this respect, it will be important to continue to study this blazar: 
we expect that an increase of $\theta$, after the very low value reached 
in the past years, would imply an increasing $\beta_{app}$ and consequently 
a larger $\langle|\dot{m}|\rangle$ (see also \citealt{Nes05}).

This consideration supports the importance of a multifrequency approach to 
distinguish geometrical from physical effects affecting the emission 
properties of BL Lac objects.
\citet{MasMan04}
pointed out how the study of both optical long-term variability and VLBI 
imaging can be useful for the understanding of geometrical and structural 
changes of the synchrotron radiation in jets of BL Lacs. 
Now we suggest the possibility that the mean properties of INOV,
say for instance $\langle |\dot{m}| \rangle $ or $ \tau_{Fm}$,
may also change on such long time scales and could be related
to the kinematics of the jet derived from VLBI imaging.
This kind of work requires the acquisition and storage of a large amount
of INOV observations for a sample of BL Lac objects, covering time intervals
of several decades. Such a great observational effort can be performed 
only with the collaboration of several groups, possibly working with
automatic/robotic small aperture telescopes, and with the creation of 
homogeneous and well organized databases.

\begin{acknowledgements}
The authors are grateful to Gino Tosti and Andrea Tramacere for fruitful
discussions.
This work was partially supported by Universit\`a di Roma La Sapienza. 
We also aknowledge the financial support Italian MIUR (Ministero dell' 
Istruzione Universit\`a e Ricerca) under the grant Cofin 2001/028773,
2002/024413 and 2003/027534. 

\end{acknowledgements}

\begin{longtable}{lrrrrccccc}
\caption{ \label{tab:logobs} Log of INOV observations of \sbz. 
(Total number of frames: 10,675).} \\
\hline \hline
Date & JD
 & Start time  & Duration & Frames & Filter & Mean mag. & $\Delta$Comp. rms & Tel. & Intervals
\\
 &($-2449000$)& UT (hh:mm)  & (hours)  &             &        &           &   \\
\hline
\endfirsthead
\caption{continued.}\\
\hline\hline
Date & JD
 & Start time  & Duration & Frames & Filter & Mean mag. & $\Delta$Comp. rms & Tel. & Intervals
\\
 &($-2449000$)& UT (hh:mm)  & (hours)  &             &        &           &   \\
\hline
\endhead
\hline
\endfoot
1996/11/12 & 1400 & 20:59 &  6.87 &  88 & B & 14.31 & 0.016 & V  & 2 \\
1997/03/03 *& 1511 & 20:00 &  6.58 &  49 & B & 14.81 & 0.016 & V  & 1 \\
1997/03/12 & 1520 & 19:11 &  6.17 &  34 & B & 14.76 & 0.008 & V  & 2 \\
1998/11/07 & 2125 & 18:29 &  9.38 & 101 & R & 13.80 & 0.008 & G  & 2 \\
1998/12/17 & 2165 & 18:13 & 10.53 & 108 & R & 14.13 & 0.008 & G  & 2 \\
1998/12/18 & 2166 & 16:49 &  5.73 &  69 & R & 14.23 & 0.006 & G  & 1 \\
1999/01/05 & 2184 & 17:22 &  8.62 &  96 & R & 14.05 & 0.009 & G  & 1 \\
1999/03/13 & 2251 & 18:51 &  8.93 & 113 & I & 13.29 & 0.007 & G  & 2 \\
1999/03/14 & 2252 & 18:59 &  3.63 &  85 & I & 12.91 & 0.007 & M  & 1 \\
1999/11/26 & 2509 & 17:55 & 11.34 & 133 & V & 14.57 & 0.018 & G  & 1 \\
1999/11/27 & 2510 & 19:01 & 10.01 & 128 & V & 14.72 & 0.013 & G  & 3 \\
1999/12/07 & 2520 & 18:02 &  6.72 &  71 & V & 14.53 & 0.009 & G  & 1 \\  
2000/01/02 & 2546 & 18:34 &  9.20 & 126 & V & 13.91 & 0.007 & G  & 1 \\
2000/01/12 & 2556 & 18:09 & 11.40 & 138 & V & 13.41 & 0.007 & G  & 2 \\
2000/01/15 & 2559 & 18:20 &  9.36 & 121 & V & 13.99 & 0.008 & G  & 2 \\
2000/01/16 & 2560 & 19:46 &  9.12 & 103 & V & 14.06 & 0.013 & G  & 1 \\
2000/01/19 & 2563 & 18:34 & 10.92 &  89 & V & 14.31 & 0.014 & G  & 2 \\
2000/01/25 & 2569 & 18:18 & 10.74 & 150 & V & 14.01 & 0.009 & G  & 4 \\
2000/02/20 * & 2595 & 20:51 &  6.00 &  67 & V & 14.41 & 0.011 & M  & 1 \\
2000/03/05 & 2609 & 19:22 &  8.37 & 101 & V & 14.69 & 0.010 & G  & 2 \\
2000/03/06 & 2610 & 18:39 &  6.04 &  71 & V & 14.41 & 0.008 & G  & 1 \\
2000/03/20 & 2624 & 18:24 &  9.37 & 109 & R & 14.15 & 0.011 & G  & 3 \\
2000/03/21 & 2625 & 19:03 &  9.05 & 106 & R & 14.13 & 0.008 & G  & 2 \\
2000/04/21 & 2656 & 19:31 &  4.61 &  62 & V & 14.17 & 0.013 & G  & 1 \\
2000/04/25 & 2660 & 19:43 &  7.12 &  86 & V & 14.30 & 0.008 & G  & 2 \\
2000/10/23 & 2841 & 18:05 &  8.88 & 130 & R & 12.77 & 0.007 & G  & 2 \\
2000/10/27 & 2845 & 18:40 &  7.72 & 111 & V & 13.09 & 0.009 & G  & 2 \\
2000/10/28 & 2846 & 18:22 &  7.01 &  97 & R & 12.67 & 0.023 & G  & 3 \\
2000/12/04 & 2883 & 16:50 &  9.61 & 117 & R & 12.87 & 0.009 & G  & 3 \\
2000/12/18 & 2897 & 17:38 &  9.36 &  83 & R & 13.85 & 0.011 & G  & 2 \\
2000/12/31 * & 2910 & 18:35 &  6.58 &  92 & R & 13.53 & 0.013 & G  & 1 \\
2001/01/14 & 2924 & 17:35 & 10.64 & 146 & R & 13.07 & 0.012 & G  & 3 \\
2001/01/21 * & 2931 & 19:27 &  7.70 &  98 & R & 13.28 & 0.009 & G  & 1 \\
2001/02/10 & 2951 & 18:44 &  9.72 & 127 & R & 13.19 & 0.009 & G  & 2 \\
2001/02/11 & 2952 & 18:33 &  9.01 & 129 & R & 13.30 & 0.006 & G  & 3 \\
2001/02/14 & 2955 & 17:45 & 10.55 & 146 & R & 13.42 & 0.006 & G  & 3 \\
2001/02/20 & 2961 & 17:45 &  6.11 &  75 & R & 13.30 & 0.007 & G  & 2 \\
2001/02/26 & 2967 & 18:33 & 10.00 & 146 & R & 13.24 & 0.009 & G  & 4 \\
2001/04/14 * & 3014 & 19:24 &  5.38 &  74 & R & 13.65 & 0.009 & G  & 1 \\
2001/04/17 & 3017 & 20:47 &  6.46 &  84 & R & 13.53 & 0.006 & G  & 1 \\
2001/04/28 & 3028 & 19:57 &  7.26 &  94 & R & 13.21 & 0.008 & G  & 2 \\
2001/08/19 * & 3141 & 20:08 &  5.58 &  67 & R & 12.99 & 0.015 & G  & 1 \\
2001/08/23 & 3145 & 21:12 &  6.12 &  82 & R & 12.99 & 0.008 & G  & 2 \\
2001/08/24 * & 3146 & 20:31 &  7.06 &  99 & R & 13.09 & 0.010 & G  & 1 \\
2001/08/26 & 3148 & 21:07 &  6.39 &  92 & R & 13.15 & 0.006 & G  & 3 \\
2001/09/03 & 3156 & 19:31 &  5.50 &  78 & R & 13.25 & 0.010 & G  & 1 \\  
2001/09/06 & 3159 & 19:58 &  7.10 &  95 & R & 13.12 & 0.007 & G  & 1 \\
2001/09/10 & 3163 & 19:08 &  8.53 & 122 & R & 13.26 & 0.009 & G  & 3 \\
2001/09/12 & 3165 & 20:22 &  7.52 & 108 & R & 13.09 & 0.011 & G  & 3 \\
2001/10/11 & 3194 & 19:47 &  7.69 & 108 & R & 12.82 & 0.007 & G  & 2 \\
2001/10/12 & 3195 & 17:59 & 10.49 & 141 & I & 12.44 & 0.009 & G  & 2 \\
2001/10/18 & 3201 & 18:32 &  3.88 &  54 & R & 12.90 & 0.006 & G  & 1 \\
2001/11/03 & 3217 & 18:07 & 10.04 & 139 & R & 13.01 & 0.007 & G  & 2 \\
2001/11/04 & 3218 & 21:58 &  5.85 &  69 & B & 13.90 & 0.007 & G  & 2 \\
2001/11/04 & 3218 & 17:25 &  3.95 &  45 & R & 13.22 & 0.008 & G  & 1 \\
2001/11/16 * & 3230 & 17:23 &  9.81 &  97 & R & 13.48 & 0.011 & G  & 1 \\
2002/01/25 & 3300 & 17:18 &  6.91 &  93 & R & 13.68 & 0.011 & G  & 2 \\
2002/02/01 & 3307 & 19:04 &  8.19 & 114 & R & 13.70 & 0.005 & G  & 2 \\
2002/02/02 & 3308 & 17:48 & 11.07 & 156 & R & 13.68 & 0.006 & G  & 2 \\
2002/02/03 & 3309 & 18:13 &  9.46 & 129 & R & 13.84 & 0.007 & G  & 2 \\
2002/03/10 & 3344 & 19:41 &  8.81 & 116 & R & 13.12 & 0.012 & G  & 2 \\
2002/03/11 & 3345 & 18:45 &  9.83 & 125 & R & 13.28 & 0.006 & G  & 2 \\
2002/03/13 & 3347 & 18:59 &  8.77 & 125 & R & 13.09 & 0.006 & G  & 3 \\
2002/03/15 & 3349 & 19:12 &  9.23 & 130 & R & 13.22 & 0.006 & G  & 2 \\
2002/03/20 & 3354 & 18:43 &  9.32 & 126 & R & 13.27 & 0.008 & G  & 2 \\
2002/03/25 & 3359 & 18:51 &  9.36 & 123 & R & 13.08 & 0.010 & G  & 4 \\
2002/03/27 & 3361 & 18:42 &  6.38 &  89 & B & 13.91 & 0.013 & G  & 2 \\
2002/04/01 & 3366 & 18:50 &  8.89 & 122 & R & 13.21 & 0.008 & G  & 2 \\
2002/04/21 & 3386 & 19:04 &  7.27 &  95 & R & 12.72 & 0.014 & G  & 3 \\
2002/04/22 & 3387 & 19:30 &  7.68 & 107 & R & 12.69 & 0.011 & G  & 3 \\
2002/05/01 & 3396 & 19:24 &  7.31 & 101 & R & 12.90 & 0.007 & G  & 2 \\
2002/05/13 & 3408 & 19:35 &  5.24 &  63 & R & 12.93 & 0.013 & G  & 1 \\
2002/05/14 & 3409 & 19:46 &  7.07 &  95 & R & 13.37 & 0.007 & G  & 1 \\
2002/05/17 & 3412 & 21:07 &  5.67 &  82 & R & 13.16 & 0.009 & G  & 1 \\
2002/05/29 & 3424 & 20:43 &  5.78 &  71 & R & 13.84 & 0.010 & G  & 1 \\
2002/05/30 & 3425 & 20:26 &  5.92 &  65 & R & 13.61 & 0.012 & G  & 1 \\
2002/09/30 & 3548 & 18:34 &  5.52 &  70 & R & 13.82 & 0.008 & G  & 2 \\
2002/10/28 & 3576 & 19:19 &  8.37 & 110 & R & 13.93 & 0.008 & G  & 2 \\
2002/10/29 & 3577 & 18:02 &  9.66 & 129 & R & 13.96 & 0.008 & G  & 2 \\
2002/11/04 & 3583 & 18:57 & 10.01 & 133 & R & 13.46 & 0.006 & G  & 2 \\
2002/11/06 & 3585 & 18:49 &  5.77 &  67 & R & 13.24 & 0.006 & G  & 2 \\
2002/12/29 & 3638 & 17:30 & 12.25 & 175 & R & 13.17 & 0.011 & V  & 1 \\
2003/01/18 * & 3658 & 18:20 &  9.60 & 132 & R & 13.53 & 0.009 & G  & 1 \\
2003/01/26 & 3666 & 18:43 & 10.95 & 153 & R & 13.20 & 0.007 & G  & 1 \\
2003/01/27 & 3667 & 19:22 &  7.30 &  99 & R & 13.15 & 0.009 & G  & 1 \\
2003/02/01 & 3672 & 18:18 & 10.95 & 156 & R & 13.19 & 0.009 & G  & 1 \\
2003/02/02 & 3673 & 19:16 &  7.97 & 104 & R & 13.16 & 0.006 & G  & 3 \\
2003/02/09 & 3680 & 18:26 & 10.72 & 138 & R & 13.36 & 0.005 & G  & 2 \\
2003/02/10 & 3681 & 18:24 & 11.03 & 149 & R & 13.33 & 0.006 & G  & 3 \\
2003/02/11 & 3682 & 18:53 & 10.16 & 144 & R & 13.57 & 0.007 & G  & 1 \\
2003/02/17 & 3688 & 17:56 &  5.19 &  69 & R & 13.44 & 0.013 & G  & 2 \\
2003/02/18 & 3689 & 18:20 & 10.92 & 159 & R & 13.34 & 0.009 & G  & 3 \\
2003/02/24 & 3695 & 19:31 &  5.91 &  85 & R & 13.34 & 0.004 & G  & 4 \\
2003/02/25 & 3696 & 18:10 & 11.00 & 157 & R & 13.03 & 0.006 & G  & 5 \\
2003/02/26 & 3697 & 18:16 & 10.58 &  75 & I & 12.57 & 0.007 & G  & 2 \\  
2003/03/04 & 3703 & 19:02 &  9.40 & 138 & R & 13.06 & 0.005 & G  & 2 \\
2003/03/07 & 3706 & 20:19 &  7.46 &  49 & R & 13.32 & 0.024 & G  & 1 \\
2003/03/13 & 3712 & 18:20 & 10.30 &  81 & I & 12.24 & 0.012 & G  & 3 \\
2003/03/18 & 3717 & 18:59 &  8.63 & 115 & R & 12.97 & 0.012 & G  & 5 \\
2003/03/19 & 3718 & 18:25 & 10.43 &  64 & I & 12.51 & 0.008 & G  & 1 \\
2003/03/22 & 3721 & 19:34 &  8.68 & 123 & R & 12.56 & 0.009 & G  & 4 \\
2003/03/23 & 3722 & 19:06 &  9.16 & 125 & R & 12.35 & 0.008 & G  & 4 \\
\end{longtable}

\appendix
\section{} \label{sec:App}
In this Appendix we report a brief description of the entire data set
of our photometric observations to study the INOV of \sbz.
All the data are organised in Table A1 as follows: each light curve
is reported in three columns containing UT time in hours (column 1),
source magnitude (column 2) and its error (column 3), derived as
written in Sect.~\ref{sec:ObsDR}; each light curve is identified by
a title with the date (DD-MM-YYYY) and the used filter.

\end{document}